\documentstyle[12pt,draft]{article}

\begin{document}

\title{SIMULATED TEMPERING: A NEW MONTECARLO SCHEME}

\author{
Enzo MARINARI$^{(a),(b)}$ and Giorgio PARISI$^{(c)}$\\[.5em]
Dipartimento di Fisica,\\
Universit\`a di Roma {\it Tor Vergata},\\
Via della Ricerca Scientifica,\\
00173 Roma, Italy\\
and Infn, Sezione di Roma {\it Tor Vergata}\\
$(a)$  and Physics Department,\\
 Syracuse University,\\
 Syracuse, N.Y. 13244, U.S.A\\
$(b)$ {\footnotesize MARINARI@ROMA2.INFN.IT}\\
$(c)$ {\footnotesize PARISI@ROMA2.INFN.IT}
}

\date{February 20, 1992}

\maketitle

\begin{abstract}

We propose a new global optimization method ({\em Simulated Tempering}) for
simulating effectively a system with a rough free energy landscape (i.e. many
coexisting states) at finite non-zero temperature. This method is related to
simulated annealing, but here the temperature becomes a dynamic variable,  and
the system is always kept at equilibrium.  We analyze the method on the Random
Field Ising Model, and we find a dramatic improvement  over conventional
Metropolis and cluster methods. We analyze and discuss the conditions under
which the method has optimal performances.

\end{abstract}

\vfill
{\hfill {\bf ROM2F-92-06, SCCS 241, hep-lat/9205018} }
\vfill

\newpage

Simulated annealing is an efficient heuristic method which is used to
find the absolute minimum of functions with many local minima: it has
been introduced independently in the framework of
the Monte Carlo approach for discrete variables in ref.~\cite{ANNMC},
and in the framework of stochastic
differential equations (of Langevin type) for continuous variables in
ref.~\cite{ANNCON}.

The essence of the method consists of the following. Let us suppose
that we are interested in finding the minimum of a function $H(X)$,
where $X$ denotes an element of the configuration space (which has
dimension $N$, where $N$ is often a very high number).
In most cases we do not know any method which can guarantee
to find the minimum of $H(X)$ with a computational effort that does not
increase more than polynomially in $N$.
In these cases one can try
as a first guess a random search starting from a random configuration
and minimizing $H(X)$ with a steepest descent algorithm.
If the number
of local minima increases as $e^{\gamma N}$, with $\gamma$ different
from zero, it often happens that this method also takes an exponentially
large number of trials (i.e.  $e^{\delta N}$, with in general
$\delta < \gamma$).

In the simulated annealing method one considers a $\beta$-dependent
algorithm which asymptotically generates the configurations $X$ with
Gibbs's probability distribution, i.e.  $e^{- \beta H(X)}$; for
definiteness we can consider the case of Monte Carlo steps.
Simulations at increasing values of $\beta$ are done (eventually at
$\beta = \infty$).
Each time $\beta$ is changed the system is driven out of equilibrium,
but that does not matter since eventually we are interested in the
$\beta=\infty$ result.

In general the simulated annealing method does not have any reason to converge
to the exact result, i.e. to provide the minimum of $H(X)$. Only if we do an
asymptotically  large number of simulated annealing runs, or if the values of
$\beta$ are changed by an infinitesimal amount at each step and an infinite
amount of Monte Carlo steps are done at each value of $\beta$, the simulated
annealing method will  converge to the exact result and will find the minimum
of $H$. But the convergence is guaranteed only if we asymptotically invest an
infinite amount of computer time. If a reasonable annealing scaling is used
($\beta$ is changed by a non-zero amount and only a finite number of Monte
Carlo cycles are done at a given value of $\beta$) we have no reason to
believe that this procedure ends up in the global minimum; indeed in the
extreme case in which $\beta$ takes only two values ($0$ and $\infty$) we find
the same  result as the random search algorithm we have described before.

The simulated annealing algorithm can however be used as an heuristic
predictor for the global minimum: one can compare the values of the
energy after many simulated annealing runs and if the probability of
ending with the global minimum is not too small, the simulated
annealing turns out to be a rather efficient algorithm. Let us note that  this
efficiency depends a lot on the shape of the phase space: if the absolute
minimum has a small basin of attraction, and is separated from the large local
minima by very high barriers, simulated annealing does not have any reason to
be a good algorithm.

Unfortunately if we want to extend the algorithm to finite temperature we are
very soon in deep trouble.  Indeed if we stop our simulations at a given value
of $\beta < \infty$, the one we want to use to evaluate observables, different
runs will give different results (if $\beta$ is sufficiently large).  In this
case we cannot just select the runs which produce the configurations with lower
energy: at $T \neq 0$ we have to minimize the free energy $F$ and not the
energy. Estimating the entropic contribution is a non-trivial task, and makes a
straightforward generalization of the simulated annealing impossible. This
problem is very severe in cases like spin glasses$~{\cite{SG}}$ or
hetero-polymers  folding$~{\cite{IMP}}$ (maybe also peptides$~{\cite{FUKU}}$)
in which there are more than one equilibrium state and we are actually
interested in knowing the relative weight which the different equilibrium
states carry in the partition function.

The method we propose in this note is meant to bypass these difficulties,
and to constitute a viable scheme to minimize free energy in an effective way.
It can be regarded as a very efficient global optimization scheme.
The basic idea of the {\em Simulated Tempering} method consists of changing
the temperature while remaining at equilibrium: this is in contrast with
the simulated annealing method, where every change of the temperature
drives the system out of equilibrium. This can be achieved by enlarging the
configuration space of the system in the following way.

We define a large configuration space, which is characterized by
the variables $X$ (the original configuration space)
and by a new variable $m$,
which can takes $M$ values  ($ m = 1 \dots M$).
The probability distribution $P(X,m)$ will be chosen to be

\begin{equation}
  P(X,m) \propto e^{-H(X,m)}\ ,
  \protect\label{EPRO}
\end{equation}

where we have absorbed the factor $\beta$ in the definition of the
Hamiltonian. We choose

\begin{equation}
  H(X,m) \equiv \beta_m H(X) - g_m\ .
  \protect\label{EHAM}
\end{equation}

Here the $\beta_m$ and the $g_m$ can take arbitrary values we assign a
priori. The $g_m$ will be a priori assigned
constants, and the $\beta_m$ will be
dynamical variables which will be allowed to span a set of values given a
priori. For simplicity we can assume that the $\beta_m$ are ordered.

It is evident that the probability distribution induced by the
Hamiltonian \ref{EHAM}, restricted to the subspace at fixed $m$, is
the usual Gibbs distribution for $\beta = \beta_m$. On the other hand
the probability of having a given value of $m$ is simply given by

\begin{equation}
  P_m \propto Z_m e^{g_m} \equiv e^{ -(\beta_m f_m + g_m)} \ ,
  \protect\label{EPRM}
\end{equation}

where the $Z_m$ are the partition functions at given $\beta_m$ (i.e.
$Z_m \equiv Z(\beta_m)$) and the $f_m$ are the corresponding free
energies.

If we make the choice

\begin{equation}
  g_m = \beta_m f_m \ ,
  \protect\label{EGM}
\end{equation}

then all the $P_m$ become equal.

If our target is to do a simulation at a given value of $\beta$, we
can take $\beta_{\tilde{m}} = \beta$ and with the choice in
eq.~\ref{EGM} we can perform a Monte Carlo simulation in which we also
allow the change of $m$ by $1$ unit. In this case the system will
be with a probability $\frac{1}{\tilde{m}}$ at $m=\tilde{m}$. Only a
fraction $\frac{1}{\tilde{m}}$ of the events will be interesting
for measuring directly expectation values at $\beta$ (if the use of an
histogram reconstruction makes also the other $\beta$ values very useful).
The frequent visits of the system to
lower values of $\beta_m$ will make it decorrelate much faster.
Indeed at lower $\beta$ values free energy barriers are lower, and the system
will find it much easier to jump.
Then, when it decides to cool off again, it
will be visiting, with the correct equilibrium probability, a different
minimum.

This method may be useful only if the transition from one value of $\beta_m$
to another happens with non-negligible probability. It is evident that
if the two contiguous values of $\beta$ are too different the probability of
accepting a change will be rather small, and that, on the contrary, if they
are too similar they will not help in decorrelating.

Let us try to compute the probability for going from $\beta_m$ to
$\beta_{m+1} \equiv \beta_m  + \delta $.  If we try to modify $\beta$,
the variation of the Hamiltonian is given by

\begin{equation}
   \Delta H = E \  \delta - ( g_{m+1} - g_m)\ ,
  \protect\label{EDH}
\end{equation}

where $E$ is the instantaneous value of the energy $H(X)$. On the
other hand we have that $g_{m+1} -g_m$ is given by the value of the
energy for some $\beta$ in between $\beta_m$ and $\beta_{m+1}$. More
precisely

\begin{equation}
  g_{m+1} -g_m = E_m \  \delta + \frac{1}{2} C_m \delta^2 +O(\delta^3)\ ,
  \protect\label{EDG}
\end{equation}

where $E_m$ is $E(\beta_m)$ ($E(\beta)$ is the expectation value of
$H(X)$ as function of $\beta$) and $C_m = \frac{dE}{\beta_m}$.
If we assume that $E$ is very close to $E_m$ the variation $\Delta H$
will be not too large under the condition that

\begin{equation}
  C_m \delta^2 = O(1)\ .
 \protect\label{ECON}
\end{equation}

One should also consider that there are thermal fluctuations in the
value of the energy which are of order of $C_m$. Condition \ref{ECON} is
equivalent to requiring that there is a non-negligible overlap in the
values of the energy computed at contiguous values of $\beta_m$.

In the usual thermodynamic limit the energy is a quantity of order $N$
and condition \ref{ECON} requires that $\delta$ is of order
$N^{-\frac{1}{2}}$, which is not a very demanding condition. The main
difficulty in the method is the required tuning in the choice of the
$g_m$. Indeed if one takes for the $\beta_m$ an unreasonable value, the
simulation could get trapped at a given value of $\beta_m$. In this respect it
is interesting to note that we are not introducing any systematic bias.
One can also think
about the possibility of performing an iterative procedure in which
the values of the $g_m$'s are adjusted during the simulation,
but we will see that already with the naive choice we are using one gets very
impressive results.

We have applied the {\em Simulated Tempering} method to the {\em Random
Field Ising Model} (RFIM), which has many features that are very
relevant to our case. It has a rough landscape, and the symmetry of
the $+$ and the $-$ state of the pure Ising model is broken by the
random magnetic field.  This is not a trivial symmetry any more, and
the flips from the $+$ to the $-$ sector (and back) is an essential
part of the dynamics.  The state oriented in the $+$ direction and the
one oriented in the $-$ direction, which macroscopically are very
similar, from a microscopic point of view are completely different.
The transition from the favoured state (which is selected by the
specific realization of the magnetic field) to the suppressed one is a
rare event.

For the RFIM an extension of the cluster update
method$^{\cite{SW,SOKAL}}$ does not give any improvement
over the local classical
Metropolis method$^{\cite{GUMAPA}}$. The system undergoes the usual
pathology of freezing already at $T > T_c$, and the spins form a large
cluster. In no way does the cluster method help in this case, for example,
to tunnel from a $+$ to a $-$ state.

We have implemented the {\em Simulated Tempering} by proposing one
$\beta$ update at the end of each sweep of the lattice spins. The
computational time required to compute the $\beta$ update is
negligible.

Let us anticipate our results: as we will show in some detail the {\em
Simulated Tempering} method helps a lot. In our test, correlation times
for observable quantities which are not sensitive to the magnetization
decrease by a factor of $6$ as compared to the Metropolis and the
cluster method. As far as the estimate of the magnetization is concerned,
the method changes the picture dramatically, allowing tunneling where
the Metropolis method is trapped in a single state, and correcting, in
some cases, wrong estimates given by the Metropolis method.

The lattice Hamiltonian is

\begin{equation}
  H = - \sum_{<i,j>} \sigma_i \sigma_j + \sum_i h_i \sigma_i\ ,
\end{equation}

where sites $i$ live on a $3$ dimensional lattice of volume $V$, the
sum runs over all first neighbors of the lattice, the spins $\sigma_i$
can take values $\pm 1$, and the site random fields $h_i$ take values

\begin{equation}
  h_i = |h| {\theta}_i\ ,
\end{equation}
where the $\theta_i$ take the values $\pm 1$ with probability $\frac{1}{2}$.

We have taken in our simulations $V=10^3$ and $|h|=1$. We have worked
with a given realization of the random magnetic field. In order to
characterize the system in fig.~$1$ we show the specific heat, and in
fig.~$2$ the magnetic susceptibility (as defined, on the finite
lattice, from the fluctuations of $|m|$). The $3$ points with errors
are from $3$ runs done by using the cluster algorithm, while the
dotted, dashed and dot-dashed lines are done by using the
reconstruction method proposed in ref.~\cite{FETAL}. The continuous
line uses the method of ref.~\cite{FETAL} by patching the $3$ data
points (for details see also ref. \cite{REVIEW}).
The reconstruction is very reliable.

We have analyzed the measured observables by means of a binning
procedure, obtaining an asymptotic estimate for the errors. We have
also focused our analysis on the study of $\tau^{int}$, which is the
relevant quantity related to the true error over measured observables.
Following ref.~\cite{WOLFF} we use an improved estimator for
$\tau^{int}$, taking in account the remainder $R$:

\begin{equation}
  \tau^{int}_{O}(W) = \frac{1}{2} + \sum_{t=1}^{W-1} \rho_{O}(t) + R_{O}(W)\ ,
\end{equation}

where

\begin{equation}
  R_O(W) \equiv \frac{\rho_O(W)}{1-\frac{\rho_O(W)}{\rho_O(W-1)}} \ .
\end{equation}

We have taken $W$ up to $20$.

The errors on $\tau_{int}$ are, when we quote an asymptotic estimate
for them, always of the order of $1$ on the last digit. We have also
monitored that $\tau_{exp}$
gives consistent results (we do not quote it here since it is always noisier
than $\tau_{int}$).

\begin{table}
\begin{tabular}{||c||c|c|c|c|c||} \hline
$\beta$     & $E_T$      & $\tau^{int}_{E_T}$ & $m$       & $\tau^{int}_{m}$ &
$ N_{iter} 10^{-3}$ \\ \hline\hline
$.24$ (MC)  & $1.1980(18)$ & $10$             & -.161[12] & $[70]$           &
$200$               \\ \hline
$.24$ (CL)  & $1.2059(22)$ & $14$             & -.180[10] & $[90]$           &
$200$               \\ \hline
$.24$ (B)   & $1.2045(19)$ &  $6$             & -.187(10) & $60$             &
$145$               \\ \hline
$.24$ (E)   & $1.2025(13)$ &  $3.7$           & -.159(10) & $40$             &
$160$               \\ \hline
$.24$ (F)   & $1.2015(11)$ &  $5.5$           & -.175(5)  & $32$             &
$290$               \\ \hline\hline
$.25$ (MC)  & $1.5286(15)$ & $7$              & -.37[6] & $[700]$            &
$200$               \\ \hline
$.25$ (CL)  & $1.5252(25)$ & $11$              & -.32[4] & $[660]$           &
$200$               \\ \hline
$.25$ (B)   & $1.5311(10)$ & $3.9$              & -.363(15) & $150$          &
$297$               \\ \hline
$.25$ (C)   & $1.5303(12)$ & $4.8$              & -.351(11) & $70$           &
$226$               \\ \hline
$.25$ (D)   & $1.5299(9)$ & $3.5$              & -.350(20) & $[370]$         &
$300$               \\ \hline
$.25$ (E)   & $1.5279(8)$ & $2.4$              & -.320(12) & $105$           &
$301$               \\ \hline
$.25$ (F)   & $1.5281(8)$ & $3.3$              & -.352(9) & $52$             &
$290$               \\ \hline\hline
$.255$ (MC) & $1.6723(12)$ & $9$              & -.35(13) & $[6000]$          &
$200$               \\ \hline
$.255$ (D)  & $1.6723(8)$ & $1.5$              & -.414(22) & $180$           &
$151$               \\ \hline
$.255$ (E)  & $1.6718(6)$ & $1.8$           & -.382(13) & $108$           &
$301$               \\ \hline\hline
$.26$  (MC) & $1.7954(8)$ & $2.8$           & -.7016(3) & $3.8$          &
$200$               \\ \hline
$.26$  (CL) & $1.7942(11)$ & $7.6$           & -.53[5] & $[1000]$          &
$200$               \\ \hline
$.26$  (B)  & $1.7925(7)$ & $1.6$           & -.476[18] & $[81]$          &
$158$               \\ \hline
$.26$  (E)  & $1.7924(6)$ & $1.15$           & -.433(13) & $52$          &
$150$               \\ \hline
$.26$  (F)  & $1.7928(5)$ & $1.75$           & -.473(10) & $64$          &
$307$               \\ \hline\hline
\end{tabular}

\protect\caption{Thermal energy, magnetization and related integrated
autocorrelation times.
Errors are in round brackets $()$. When in square brackets, $[]$, error
and $\tau^{int}$ estimates are not asymptotic. The value for $m$ given
by the Metropolis method (MC) at $\beta=.26$ is wrong.
\protect\label{TAB1} }

\end{table}

\begin{table}

\begin{tabular}{||c||c|c|c|c|c||} \hline

Run &$\beta_1,n_1$&$\beta_2,n_2$&
$\beta_3,n_3$&$\beta_4,n_4$&$\beta_5,n_5$\\\hline\hline
$B$ &$.24$, $145$  &$.25$, $297$  &$.26$, $158$  &
             &             \\\hline
$C$ &$.23$, $206$  &$.25$, $226$  &$.27$, $167$  &
             &             \\\hline
$D$ &$.245$, $148$ &$.25$, $301$  &$.255$, $151$ &
             &             \\\hline
$E$ &$.24$, $149$  &$.245$, $300$ &$.25$, $301$  &
$.255$, $301$ & $.26$, $150$ \\\hline
$F$ &$.23$, $159$  &$.24$, $290$  &$.25$, $290$  &
$.26$, $306$  & $.27$, $155$ \\\hline\hline

\end{tabular}

\protect\caption{
\protect\label{TAB2}
$\beta$ values allowed in each of our {\em Simulated Tempering} runs,
and number of iterations (in units of $10^3$) the system spent at each
$\beta$ value. For historical reasons we label the runs with the
capital letters $B$, $C$, $D$, $E$, $F$.  }
\end{table}

In table \ref{TAB1} we give two of the measured observables: the
thermal part of the energy (i.e. the expectation value of $\sigma_i
\sigma_j$), $E_T$, and the magnetization $m$. $E_T$ has a behavior
typical of the quantities that are $Z_2$ symmetric.  The lines called
(MC) and (CL) give information about the runs we have done with the
classical Metropolis method and with the cluster algorithm. These runs
have been used (together with more MC runs at other $\beta$ values) to
get a preliminary estimate of the system energy and to determine the
values of the $g_m$. It is in no way necessary to get, for estimating
the $g_m$, more than a rough estimate of the $E_m$, and in a practical
application of the method the preliminary MC runs can be very short.
It is possible to determine directly the values of the $e^{-f_n}$, by
using the energy histograms taken in the preliminary runs. Although we
stress that this possibility exists, we do not think
that it could dramatically increase
the efficiency of
the method. When, in table
\ref{TAB1}, we put errors and $\tau_{int}$ in square brackets we mean that
we did not get an asymptotic estimate.  Let us also note now that the
MC run at $\beta=.26$ gets a {\em wrong expectation value} for $m$. In
this case the standard Metropolis does not produce any tunneling
event, and always stays in the $-$ phase.

In table \ref{TAB2} we give details about our {\em Simulated Tempering}
runs. We have tried different combinations, allowing the system to
take $3$ or $5$ $\beta$ values, always centred around $\beta = .25$.
In table \ref{TAB2} we check the performance of our method at the
different $\beta$ values we have allowed in the different simulations.
The choice of the $\beta$ values has been dictated, as we have
discussed before, by the requirement of having a non-negligible
overlap in the energy histograms of the preliminary MC runs. Runs $D$
and $E$ have a very small $\delta$ value, and a high acceptance
factor for a $\beta$ update, of $\simeq 70 \%$. Runs $B$ and $F$ have
a medium $\delta$, and a $\beta$ acceptance factor of $40-50 \%$. Run
$C$ has a higher $\delta$ value and a very low acceptance factor for
the $\beta$ update, $10-15 \%$.

In fig.~$3$ we give $\beta_m$ as a function of the computer time for
system $B$, where $3$ $\beta$ values were allowed and the $\beta$
acceptance value is close to $50 \%$.  Let us start by commenting on
the results for $m$, which are quite spectacular. At $\beta=.24$ (not
so low $T$) $\tau_m$ is higher than $O(100)$ for Metropolis and
Cluster methods, and gets down to $32$ in the $F$ run.  In general
runs with a larger $\delta$ value seem to be more effective for
improving the estimate of $m$.  Things are better and better at lower
temperatures. At $\beta=.25$ from $\tau_m > 700$ we go down to
$\tau_m=52$ in run $F$, with a gain of a factor larger than $12$. At
$\beta=.255$ from $\tau_m>6000$ we go down to $108$ in run $E$, with a
gain of a factor better than $60$. At $\beta=.26$ after $200000$ steps
the Metropolis does not succeed in getting a single tunneling event, while
our run $E$ has $\tau_m=52$. In figs.~$4a-c$ we show what happens. In
fig.~$4a$ we give the magnetization as a function of computer time for
the Metropolis method, for $200000$ steps. The system stays in the
$-$ state, with very large fluctuations which never succeed in getting a
complete flip. In fig.~$4b$ we plot $m$ for our $F$ system, only
$1000$ steps. Here the data points are at different $\beta$ values, and
it is clear that going to different $\beta$ values allows an easy
flipping. In order to make the situation clear in fig.~$4c$ we have
selected only the first $10000$ configurations, of the $F$ dynamics,
which happen to be at $\beta=.26$. The picture speaks for itself.

Also for $E_T$ there is a large gain at all $\beta$ values. One gains a factor
$3$ at $\beta=.24,.25$, a factor $6$ at $\beta=.255$, and a factor $2.5$ at
$\beta=.26$. In this case the best performances are obtained for small
$\delta$ values.

It is a pleasure for us to thank Masataka Fukugita for
interesting discussions, and Paul Coddington for a critical reading of the
manuscript. Hardware and software computer support has been given
by Infn (Roma {\em Tor Vergata}) and NPAC (Syracuse).


%
\vfill
\newpage

{\bf Figure Captions.}

\begin{description}
  \item[$1$]
   Specific heat $C_V$. Points from cluster algorithm data,
lines from hi<stogram
     reconstruction.
  \item[$2$]
	    As in fig.~1, but susceptibility $\chi$.
  \item[$3$]
     $\beta$ as a function of the computer time for the runs of
series $B$ ($3$ $\beta$ values allowed, $50 \%$ acceptance ratio).
  \item[$4a-c$]
     Magnetization $m$ as a function of computer time. In $(a)$ for
the Metropolis method at $\beta=.26$, in $(b)$ $m$ for the $F$
systems ($\beta$ is here a dynamical variable which is allowed to
take $5$ values during the course of the dynamics), in $(c)$ the
configurations of run $F$ which have $\beta=.26$.
\end{description}

\end{document}